\documentclass[12pt]{article}

\newcommand{\be}{\begin{equation}} \newcommand{\ee}{\end{equation}}
\newcommand{\ba}{\begin{eqnarray}} \newcommand{\ea}{\end{eqnarray}}
\newcommand{\bit}{\begin{itemize}} \newcommand{\eit}{\end{itemize}}
\newcommand{\ben}{\begin{enumerate}} \newcommand{\een}{\end{enumerate}}

 \newcommand{\dpa}{\partial}
\renewcommand{\Re}[1]{\mbox{\rm Re}\left(#1\right)}
\renewcommand{\Im}[1]{\mbox{\rm Im}\left(#1\right)}

\begin{document}

\hspace*{0.29\textwidth} 
 \underline{\bf http://xxx.lanl.gov/e-print/physics/0208098}\\

\begin{center}
\textbf{\Large
Some notes on ideology of waves in plasmas}
\vspace{3mm}

V.~N.~Soshnikov
\footnote{Krasnodarskaya str., 51-2-168, Moscow 109559, Russia.}
\vspace{1mm}

Plasma Physics Dept.,\\
All-Russian Institute of Scientific and Technical Information\\
of the Russian Academy of Sciences\\
(VINITI, Usievitcha 20, 125315 Moscow, Russia)
\end{center}
\vspace{-2mm}

\begin{abstract}
  Our last three papers~\cite{bib-4,bib-7,bib-8} 
 provide an occasion to make some brief notes 
 on ideology of waves in plasmas
 and to rehabilitate Vlasov prescription to calculate relevant 
 logarithmically divergent integrals in the principal value sense.
 In this approach asymptotical solutions of plasma oscillations 
 obtained with the method of Laplace transform 
 must be in the form of a sum of wave exponents
 with amplitudes being selected 
 according to self-consistent boundary physical conditions. 
 Landau damping is absent in this case by definition. 
 Boundary electrical field together with conditions of absence 
 of unphysical backward waves 
 (for boundary problem in half-infinite plasmas) 
 and of kinematical waves 
 define single-valued dependence of boundary distribution function 
 on electron velocity $\vec{v}$ 
 in the case of transversal waves 
 and on the surface break of the normal electrical field 
 in the case of longitudinal oscillations. 
 We have proposed physically more justified
 modified iteration procedure of collisional damping calculation
 and demonstrated some results of damping decrements calculations
 in a low-collision electron-ion plasma. 
 We have pointed out a possibility of natural selection 
 of boundary conditions, 
 which allows to eliminate simultaneously both backward and kinematical waves 
 as for transversal as well as for longitudinal oscillations. 
 Dispersion smearing of both longitudinal and transversal 
 high-frequency waves, 
 for which the smearing decrement  $\delta_x$ 
 is proportional to 
 $\Delta\omega/(\omega\sqrt{\omega^2-\omega_L^2})$ 
 (where $\Delta\omega$ is the frequency interval of exciting boundary field), 
 and non-Maxwellian background distribution function
 might be the main causes of waves amplitude damping 
 in collisionless plasmas imitating Landau damping.        
\end{abstract}

PACS numbers: 52.25 Dg; 52.35 Fp.

Key words: {\em plasma oscillations; plasma waves;
                Landau damping; Coulomb collisions; collision damping;
                dispersion equations; Vlasov equations; 
                plasma dielectric permittivity; kinematical waves; plasma echo;
                Liouville theorem}.


    After the basic paper of L.~Landau in 1946~\cite{bib-1},
which stated that even in collisionless Maxwellian plasmas 
electron waves are damping ("Landau damping"), 
there appeared an enormous amount of works, including all textbooks,
that use it as a conception. 
In practice this scientific trend can be considered 
as being completed without any hope 
to carry therein something essentially new.

    Landau had considered the simplest case of plane half-infinite geometry 
with the beforehand given form of a solution 
$\exp(ikx)f(t)$,
where for determination of the function $f(t)$
he had used Laplace transformation,
and had postulated 
to calculate an appearing in Laplace image $f_p$
logarithmically divergent integral 
not in the sense of the principal value 
(as it had been proposed in 1937 by Vlasov~\cite{bib-2}
 for the analogous integral appearing at solving his ``Vlasov equations") 
but according to Landau rule of by-passing poles 
of $dv_x$-integrand 
in the complex-valued plane $v_x$ 
($v_x$ is an x-component of the electron velocity,
 $x$ is a direction of the wave traveling).

    However it is well known 
(see, for instance review ~\cite{bib-3} or Eq.~(1) in~\cite{bib-4})
that asymptotical Landau solution 
in the form of a solitary plane damping wave 
$\exp(ikx-i\omega t)$ with $\omega=\omega_0-i\delta$
does not satisfy Vlasov equations 
neither at these
nor at any other real 
(if considered as a limit at tending 
imaginary parts of $\omega$, $k$ to zero, see, for instance,~\cite{bib-4}),
or complex values $\omega$ and $k$.
In spite of some paradoxes 
and hardly to be explained consequences of Landau theory 
(see, for instance,~\cite{bib-5})
it was supposed that Landau solution 
corresponds to reality and appears to be true,
so that Vlasov equations must be corrected 
by additional terms,
which are determined by Landau rule of by-passing poles 
(see, for instance,~\cite{bib-3,bib-6}). 

  Nevertheless one can formulate a problem 
of finding asymptotical solutions of Vlasov equations 
and also equations of propagating transversal electromagnetic waves 
(with the same difficulty of logarithmically divergent integrals)
without these additions,
which appear to be artificial.
So, the paradoxes of Landau solution are thought to be generated 
by Landau representation of logarithmically divergent integrals 
and, in the main, by the single-exponent form 
(suggested by Landau as well as by Vlasov,
 however with no paradoxes in the last case ) 
of an asymptotical solution.

  Such asymptotical solutions have been written 
in our last papers~\cite{bib-4,bib-7,bib-8}
proceeding from the following propositions:
\begin{enumerate}
  \item The logarithmically divergent integrals 
  appearing at Laplace transformation 
  have to be calculated in Vlasov prescription of the principal value sense. 
  \item For evaluation of principal values of integrals 
  we use a series of successful approximations of the type
  \ba
   \int\limits_{-\infty}^{\infty}
    \frac{e^{-\frac{mv^2}{2kT_e}}\cdot v_x d^3\vec v}
         {p_1+v_xp_2}
  &\equiv& -8p_2
   \int\limits_{0}^{\infty}
    \frac{e^{-\frac{mv^2}{2kT_e}}\cdot v_x^2 d^3\vec v}
         {p_1^2-v_x^2p_2^2} \simeq \nonumber\\
  &\simeq& -\left(\frac{2\pi kT_e}{m}\right)^{3/2}
             \frac{p_2\bar{v_x^2}}{p_1^2-\bar{v_x^2}p_2^2}\,.
   \label{eq-1}
  \ea
  After this the obtained asymptotical solutions 
  with the collisional damping
  take the form of a sum of exponents 
  of the type 
  $\exp(\pm ikx \pm i\omega t + x\delta)$ and 
  $\exp(\pm ikx \mp i\omega t - x\delta)$,
  $\delta>0$, 
  corresponding to double poles $p_1$, $p_2$.
  \item The given boundary condition $E(0,t)$
  (for instance, in purely boundary problem) 
  defines also, in a single-valued way, 
  the boundary (and initial) conditions for the perturbed distribution function 
  $f_1(\vec{v},0,t)$, $f_1(\vec{v},x,0)$
  according to linear integral equations 
  following from the condition of absence of kinematical waves~\cite{bib-8}.
  It guaranties the physically justified proportionality 
  of the boundary and initial distribution function
  to the boundary electric field.    
  \item Selecting free boundary parameter $F_{p_1}$, 
  being a Laplace transform of $\dpa E(x,t)/\dpa x|_{x=0}$
  one can exclude (cancel) 
  unphysical (for the case of half-infinite plasmas) backward waves
  contained in the solution in the form of residua exponential sum.
  These waves are divergent at $x\to\infty$ 
  in the case of a low-collision plasma.
  \item In the case of longitudinal waves 
  at the given boundary field $E(0,t)$ 
  the boundary self-consistent function $f_1(\vec{v},0,t)$ is determined 
  analogously from the condition of absence of kinematical waves. 
  But for elimination of the unphysical backward waves in this case 
  one must inevitably assume 
  a break in the normal constituent of electrical field $E(0,t)$ 
  at plasma surface 
  (see~\cite{bib-8} for a way to calculation its value). 
  \item In general case the solution can not be represented 
  as a single exponent even if the boundary condition $E(0,t)$
  has such a form.
  At the boundary field frequency $\omega$
  the asymptotic solution must contain at least two simultaneously existing 
  forward wave exponents $\exp(\pm i \omega t \mp i k x)$,
  corresponding to the frequencies $+\omega$ and $-\omega$.   
  \item Amplitudes of all modes with different $k_n(\omega)$
  corresponding to the given exciting field $E(0,t)$
  are strongly correlated. 
  Cancellation of kinematical and then backward waves 
  must be achieved for all modes 
  in the general asymptotical solution.
  \item Solutions of the wave equations must be real-valued 
  whenever the boundary and initial conditions are such ones.
\end{enumerate}
  Only after proceeding these procedures 
one can calculate strong relations between amplitudes 
of the different forward waves including, in the general case, 
both electron and also ion and hybrid ion-electron branches.     
    
  This allows to avoid all paradoxes at calculation of $E(x,t)$
and paradoxical tangling of the distribution function in $v_x$
with generating some strange electrical field.
    
  When all these conditions being fulfilled
one can easily construct,
using the method of two-dimensional Laplace transform
usual solutions for the poles of image $E_{p_1p_2}$,
correspondingly, 
the long before known dispersion relations 
$k(\omega)$~\cite{bib-4,bib-7,bib-8}.
Thus, for the case of collisionless plasma 
mystic Landau damping and kinematical waves are really absent by definition. 
The solution appears as a sum of exponents  
with amplitudes selected in accordance 
with boundary and physical conditions. 
    
    We also have developed the more physically justified modification 
of iteration process 
in the presence of collision terms of kinetic equation 
(low-collision electron-ion plasmas). 
By this method we have obtained damping decrements 
as for electron longitudinal waves 
as well as for two branches (low-  and high-frequency ones) 
of transversal electromagnetic waves 
with the unusual\footnote{nonlinear in concentrations of charged particles.} 
decrement for the low-frequency branch. 
    The collisional dissipative absorption 
in the high-frequency branch at $\omega>\omega_L$ is proportional to 
$1/\sqrt{1-\omega_L^2/\omega^2}$
and grows to infinity at $\omega\to\omega_L$~\cite{bib-8}. 
    
 In this connection we think that 
 Van Kampen waves (see [3]) are called 
 to compensate the erroneous consequences of Landau theory.
    
    It is highly believed that Landau damping is detected and verified experimentally. 
In this respect we should note
the relatively small number of such verifications 
and the necessary extreme delicateness of similar experiments. 
The theoretically absent Landau damping in fact
might be imitated by a series of secondary effects. 
    These ones can be:
\ben
 \item[(i)] the difference of the background distribution 
 from Maxwellian one due to electron collisions with 
 and recombinations on the walls of discharge tube; 
 \item[(ii)] the presence of longitudinal magnetic field 
 and cyclotron motion and transversal diffusion to walls; 
 \item[(iii)] effects of the method of plasma oscillations exciting;
 \item[(iv)] effects of reflecting from walls;
 \item[(v)] effects of the base electric field supporting discharge;
 \item[(vi)] growing to infinity of the Coulomb collision damping at 
  $\omega\to\omega_L+0$;
 \item[(vii)] non-harmonic composed waveform and signal dispersion 
 and its smearing in $\omega$;
  \item[(viii)] experimental requirements to electron distribution function 
  to be Maxwellian and electrons to be collisionless  are intrinsically 
  contradictory; 
  \item[(ix)] geometrical (diffraction) effects (declination from the proposed 
  one-dimensional problem with plane waves 
  since the length of the discharge tube is much more than its diameter).
\een    

    The decrement of the amplitude damping at dispersion smearing 
of a wave can be calculated according to expression 
\begin{equation}
 \label{eq-2}
  \delta_x(\omega)
  = \frac{\Delta\omega}{2\pi}
    k(\omega)
    \frac{\dpa^2\omega/\dpa k^2}
         {\left(\dpa\omega/\dpa k\right)^2}
  = \frac{\omega_L^2\Delta\omega}
         {2\pi\beta\omega\sqrt{\omega^2-\omega_L^2}}\,,
\end{equation}
where $\beta=c$ or $\beta=\sqrt{\bar{v_x^2}}$ 
for transversal or longitudinal waves, correspondingly; 
$\Delta\omega$ is spectral width of the boundary exciting field $E(0,t)$.
This smearing might be the main cause of the wave amplitude damping 
in collisionless plasma 
both for longitudinal and transversal waves
that imitates Landau damping.                                             
    
    The return to Vlasov prescription of calculating relevant integrals 
in the principal value sense 
and the proper determination and using of mutually dependent boundary 
(or initial) conditions 
in the self-consistent manner 
allow to solve and remove all paradoxes of ``Landau damping".
    
    Thus, the right natural and simple but non-traditional ideology of waves 
in plasmas reduces to combine the method of Laplace transformation 
with the proper account for boundary/initial conditions, 
and Vlasov prescription of the principal value sense of relevant integrals.

    Mathematical correctness of the represented solution 
follows from the way of its construction.
However one can analogously construct also other mathematically 
irreproachable solutions, 
for instance using ``Landau functions'',
that is with integrals in the principal value sense 
plus results of passing around the running with $p_1$ and $p_2$ 
poles $v_x=-p_1/p_2$
in the lower or upper half-plane 
(in either case of fixed $p_1^0$ with two roots 
 for the further calculated poles $p_2^0$).
 
    It appears that the only criterion to pick out a physical solution 
from the set of mathematically correct solutions 
to be only physical reasons, 
namely the definition of logarithmically divergent integrals 
and related together boundary conditions
in a such manner to avoid the paradoxes of 
divergent at $x\to\infty$ wave solutions, 
appearance of kinematical and backward waves, 
and waves not related to the boundary electrical field. 
The stability of the physical solution might be also considered
as an additional requirement.

    Namely, these and only these criterions 
justify constructing the solution in terms of functions 
with integrals in the principal value sense 
as an optimal and apparently the unique variant.

    Violation of the uniqueness and existence theorem 
is related with singularity of the Laplace image of $f_1^{(e)}$
due to logarithmic divergence of the integral in $dv_x$ 
at points $v_x=-p_1/p_2$,
where $p_1$ and $p_2$ are the running Laplace transform parameters.

    Attempts to establish functional constrains 
on the maximal value of the divergent integral 
naturally prove to be artificial and\footnote{%
since real physical values are limited by different physical processes 
which are not accounted in the original equations 
in the given formulation of the problem.} 
can be justified only by special physical considerations.

    As a conclusion, our principally new approach in the plasma wave problem 
is the prescription of self-consistent calculation 
of the lacking boundary and initial conditions 
that gives a possibility to calculate the correct relative amplitudes 
of all relevant oscillatory modes 
and to eliminate the known paradoxes. 
Here acts a simple and robust principle: 
in the self-consistent problem there must be no perturbations 
which are not induced by (and as a consequence nonlinearly related with) 
the perturbing boundary electrical field.

Transition to the real-valued  boundary condition 
$E(0,t) = E_0\cos(\omega t)$
in the asymptotical two-wave solution
$a\cdot\exp(i\omega t - i kx)+a^*\cdot\exp(-i\omega t + i kx)$ 
is equivalent to representing the solution as (cf.~\cite{bib-2})
\begin{eqnarray}
 \label{eq:3n}
 E(x,t) 
 &=& E_0' \cos\left[\omega t - kx + \varphi(k)\right]\,;\\
 \label{eq:4n} 
 f_1(v,v_x,x,t) 
 &=& F(v,v_x) \cos\left[\omega t - kx + \psi(k)\right]\,,
\end{eqnarray}
where 
 $E_0'$, $\varphi(k)$, $F(v,v_x)$, $\psi(k)$
are real-valued amplitudes and phases which have to be determined. 
The substitution of $E(x,t)$ and $f_1(v,v_x,x,t)$
into Vlasov equations at $\varphi(k)-\psi(k)=\pm\pi/2$
leads to the traditional equations
\begin{eqnarray}
 \label{eq:5n}
 F(v,v_x)
 &=& \pm \frac{|e|}{m} 
      \left(\frac{E_0'}{\omega-kv_x}\right)
       \frac{\dpa f_0(v)}{\dpa v_x}\,,\\
 \label{eq:6n} 
 1
 &=&  \frac{-4\pi e^2}{m}
       \frac{1}{k}
        \int
         \frac{\dpa f_0(v)}{\dpa v_x}
          \frac{d^3\vec{v}}{\omega-kv_x}
\end{eqnarray}
with real roots $\pm k$ of Eq.(\ref{eq:6n}) at $\omega>\omega_L$.
If Eqs.(\ref{eq:5n}) and (\ref{eq:6n}) are satisfied,
then $E(x,t)$ (\ref{eq:3n}) and $f_1(v,v_x,x,t)$ (\ref{eq:4n}) 
are asymptotical solutions of Vlasov equations. 

    The case of the totally imaginary $k$ 
($\Re{k}=0$, $\omega<\omega_L$) in collisionless plasmas 
corresponds to the total wave reflection without energy dissipation.
According to the afore-mentioned theory 
the solution reduces to a sum of terms,
which equally exponentially damp at $x\to\infty$.
As can be verified by direct substitution into Vlasov equations,
elementary general asymptotical solution has the following form
\begin{eqnarray}
 \label{eq:7n}
  E(x,t)
   &=& E_0' e^{-|k|x}
      \left[\cos(\omega t+\varphi)
         +  \cos(\omega t+\psi)
      \right]\,,\\
\label{eq:8n}
  f_1(v,v_x,x,t)
   &=& e^{-|k|x}
      \left[F(v,v_x)\cos(\omega t+\varphi)
         +  F(v,-v_x)\cos(\omega t+\psi)
      \right]\,,\\
 \label{eq:9n}
 F(v,v_x)
 &=& \mp \frac{|e|}{m} 
      \frac{\dpa f_0(v)}{\dpa v_x}
       E_0'
       \frac{\omega\pm|k|v_x}{\omega^2+|k|^2v_x^2}\,,
\end{eqnarray}
with $\varphi-\psi = \pm\pi/2$ and 
dispersion equation being defined by Eq.(\ref{eq:6n}).

In dependence on the sense of improper integral in Eq.(\ref{eq:6n}) 
one obtains different solutions $k(\omega)$,
with functions (\ref{eq:3n}) and (\ref{eq:4n}) 
being indeed the asymptotical solutions of Vlasov equations. 
From the real-valued character of the solution 
it follows directly 
the necessity to exclude in Eq.(\ref{eq:6n})
additive imaginary constituents
related to passing around the pole $v_x=\omega/k$, 
that is, 
to exclude the Landau prescription how to calculate this integral. 
From the physical symmetry of frequency and velocity of propagation
of the forward and backward waves 
with respect to interchange $k\to-k$
it follows also that one has to calculate 
the integral in Eq.(\ref{eq:6n}) in the sense of principal value. 
This becomes more evident when one uses our approximate expression 
for the principal value of this integral. 
For calculation of the phase $\varphi(k)$, 
amplitudes and initial and boundary values 
$f_1(v,v_x,x,0)$, $f_1(v,v_x,0,t)$ 
in order to take into account several modes 
$k_1(\omega)$, $k_2(\omega)$, $\ldots$ 
one ought to use the general procedure 
of Laplace transform method 
with exclusion of kinematical and backward waves. 

 Note, that solution (\ref{eq:5n}) at $v_x\to\omega/k$
tends to $\pm\infty$, 
violating the initial perturbative condition $|f_1|\ll f_0$.
However, one can assume from the physical point of view
that there occurs some saturation of $|f_1|$ growth
at periodically sign-alternating $f_1$
with limitation 
$|f_1(v_x=\omega/k\pm\varepsilon)|\leq f_0(v_x=\omega/k)$ 
at $\varepsilon\to0$,
and presumably $f_1(v_x=\omega/k)=0$
with linear dependence $f_1$ on $v_x$
in this region of unapplicability of the kinetic equation
and possible discontinuity of $\dpa f_1/\dpa v_x$
near $f_1\simeq \pm f_0$,
so, at least, there are no physical reasons 
for appearance of any imaginary part in the integral (\ref{eq:6n})
and in the function $f_1$.\footnote{%
In this case $\int|f_1|d\vec{v}\equiv|\Delta n_e|$
can be treated as the real-valued concentration of perturbed electrons
($|\Delta n_e|\leq n_e$).}
All these appear to be additional arguments in favor 
of calculation of the divergent integral in (\ref{eq:6n})
in the principal value sense. 
The case of complex $k$ in forward waves 
($\textbf{Re}\, k > 0$, collision damping) 
was considered in our previous papers 
using the general Laplace transform method. 
In this case 
for real-valued boundary (and initial) conditions and $\omega$ 
the solution is defined with a pair combination of exponents of the type
$a\cdot\exp(i\omega t - i kx)+a^*\cdot\exp(-i\omega t + i k^*x)$. 

Owing to the symmetry $k\to-k$ in isotropic medium, 
the wave number $k$ must enter into dispersion equation 
only through $k^2$. 
In collisionless plasmas asymmetry of $|f_1|$ 
relative to the neighbourhood of the point $v_x=\omega/k\pm\varepsilon$
leads to the possibility of appearing complex $k$,
which define both damping and growing waves.
Therefore this asymmetry leads to the paradoxical availability of waves
(both in forward and backward directions)
with exponential damping as well as with exponential growing
due to the fact that asymptotic solution in this case contains 
both exponents $\pm i(\omega t+kx)$ and $\pm i(\omega t-kx)$.

The fact of appearance of the indefinitely divergent integrals 
is the direct consequence of incomplete information 
in original differential equations. 
There must be additional exclusively physical (non-mathematical!) considerations
at calculation of these integrals,
since all different possible ways lead to different equally right 
(by their  Laplace transform construction) mathematical solutions 
with correspondingly different dispersion relations $k(\omega)$
and different wave amplitudes. 
But these solutions can be as physical as well as unphysical ones 
and have to be selected.

To resume, the functions (\ref{eq:3n}) and (\ref{eq:4n}),
as well as (\ref{eq:7n}) and (\ref{eq:8n}),
are evidently the very simple illustration 
of the existence of asymptotical non-damping 
or non-dissipative real-valued solutions 
for plasma wave equations.

   In the resonance region, where the condition 
$f_0(v,v_x,x,t)+f_1(v,v_x,x,t)\geq0$ is violated, 
the original, linearized or precise, equations are inapplicable. 
The divergence of $f_1$ at $v_x=\omega/k$ can not be removed 
when one uses expansion of the precise kinetic equation 
with the quadratic term 
in multiple overtones $F_n\cos[n(\omega t-kx)+\varphi_n]$ and 
$E_n\cos[n(\omega t-kx)+\psi_n]$.
Thus, the original equations, linearized or not, 
are not sufficient to obtain single-valued solution of the problem 
without introducing into the kinetic equation 
(which details Liouville theorem) some additional physical specifications.

   In the theory of ``Landau damping''~\cite{bib-1}
velocity $v_x$ can take, according to Cauchy theorem, 
some arbitrary complex values at arbitrary deformation 
of the integration contour bypassing real-valued pole $v_x=\omega/k$ 
(with $\omega>\omega_L$, $\Im k=0$) in the complex plane, 
that leads simultaneously to the appearance 
of exponentially damping and growing wave solutions
with non-zero $\pm\Im{k}$. 
According to~\cite{bib-1} the latter is related 
with the small non-zero values $\Im k$ 
resulting from the next derived quadratic dispersion equation. 
It evidences the intrinsic discrepancy and non-self-consistency 
of the Landau damping theory, since complex $k$, 
according to~\cite{bib-1}, 
is a result of solving the dispersion equation, 
contrary to the before assumed real-valued $k$ with the only one pole $\omega/k$
on the real axis $v_x$,
thus the pole bypassing in the complex plane $v_x$ 
ought \textit{e.g.} to be made now not along the half-circle 
(as it is presented in~\cite{bib-1}), 
but along the total circle~\cite{bib-9}, 
or must not be made at all, 
with not clear grounds after all, 
for taking any value of $k(\omega)$.

\section*{Conclusion}
 The uniqueness theorem for solutions of the plasma wave equations, 
both for longitudinal and transversal waves at definite boundary and initial conditions, 
is violated due to the presence of logarithmically divergent integral 
in dispersion equations. 
To avoid the appearing of exponentially divergent solution terms, 
e.~g. for considered plane waves in isotropic Maxwellian plasmas 
in the plasma half-infinite slab, 
one ought to treat this divergent integral in the principal value sense. 
The physical requirements for the solution 
select the only solution out of all mathematically possible ones. 
We have presented elementary general real-valued solutions 
of Vlasov equations for collisionless plasmas 
at real-valued boundary and initial conditions 
in the form of a trigonometric function sum 
both for non-damping plasma waves ($\Im{k}=0$, $\omega>\omega_L$)
and for the total reflection case ($\Re{k}=0$, $\omega<\omega_L$) 
with dissipationless damping (evanescence) proportional to $\exp(-|k(\omega)|x)$. 

    Finiteness of the wave solution in the neighbourhood of the point 
$v_x=\omega/k\pm\varepsilon$
results by no means from energy dissipation in the sense of contradictory
Landau damping
but just from the purely kinematical effect of the electron density limitations
with providing proper corrections to the kinetic equation.

   Existence of the general real-valued non-damping or 
dissipationless finite solutions 
of Vlasov equations is the direct proof of incorrectness of Landau damping theory.
\vspace*{9mm}

\textbf{Acknowledgements}
  The author is thankful to Dr.~A.~P.~Bakulev 
for criticism and assistance in preparing the paper in \LaTeX\ style.


\begin{thebibliography}{9}
\bibitem{bib-1}
 Landau~L.~D.,
  J. Phys. (USSR), \textbf{10} (1946) 25;\\
  JETP (USSR), \textbf{16} (1946) 574 (in Russian);\\
  Uspekhi Fiz. Nauk, \textbf{93} (1967) 527 (reprint, in Russian).
\bibitem{bib-2}
 Vlasov~A.~A.,
  JETP (USSR), \textbf{8} (1938) 291 (in Russian);\\
  Uspekhi Fiz. Nauk, \textbf{93} (1967) 444 (reprint, in Russian).
\bibitem{bib-3}
 Pavlenko~V.~N., Sitenko~A.~G., 
  "Echo-phenomena in Plasma and Plasma-like Media",
    Nauka, Moscow (1988) (in Russian).
\bibitem{bib-4}
 Soshnikov~V.~N.,
  "Damping of plasma-electron oscillations and waves 
   in low-collision electron-ion  plasmas",
  physics/0105040 (http://xxx.lanl.gov/e-print).
\bibitem{bib-5}
 Clemmow~P.~C., Dougherty~J.~P., 
  ``Electrodynamics of Particles and Plasmas'', 2-nd ed., 
   Addison-Wesley, NY (1990); (Rus. transl. Moscow, Mir, 1996).
\bibitem{bib-6}
 Alexandrov~A.~F., Bogdankevich~L.~S., Rukhadze~A.~A.,
  ``Foundations of Electrodynamics of Plasma'',
   2nd ed., Vysshaya Shkola, Moscow (1988)
    (in Russian).
\bibitem{bib-7}
 Soshnikov~V.~N.,
  "Damping of transversal plasma-electron oscillations and waves
   in low-collision electron-ion  plasmas",
  physics/0111014 (http://xxx.lanl.gov/e-print).
\bibitem{bib-8}
 Soshnikov~V.~N.,
  "Damping of electromagnetic waves 
   in low-collision electron-ion plasmas",
  physics/0205035 (http://xxx.lanl.gov/e-print).
\bibitem{bib-9}
 Alexeff~I., Rader~M.,
   Int.\ J.\ Electronics, \textbf{68} (1990) 385.
\end{thebibliography}
\end{document}